\documentclass[lettersize,journal]{IEEEtran}

\usepackage{amsmath,amssymb,amsfonts}

\usepackage{makecell}
\usepackage{xspace}
\usepackage{cancel}
\usepackage{hyperref} 
\usepackage{booktabs}

\usepackage{multirow}
\usepackage{array}
\usepackage{graphicx}
\usepackage{textcomp}
\usepackage{xcolor}
\usepackage{caption}
\usepackage{comment}
\usepackage{booktabs}
\usepackage{url}
\usepackage{subcaption}
\usepackage{flushend}

\newcommand{\imcol}{{\texttt{im2col}}}
\newcommand{\microqonv}{{\sc MicroQonv}}
\newcommand{\microqonvp}{{\sc MicroQonv+}}
\newcommand{\MicroQonv}{{\sc MicroQonv}}

\begin{document}


\title{\MicroQonv: Reshaping Convolution Tensors for Efficient Microscaling in Training and Inference}


\author{Romain Facq, Sami Ben-Ali, Olivier Sentieys\\Univ Rennes, Inria, CNRS, IRISA, Rennes, France}


\maketitle

  



\begin{abstract}
Microscaling quantization techniques are increasingly used to represent neural network parameters with 8 bits or fewer while preserving near-full precision accuracy. 
However, applying these methods efficiently in convolutional layers is not straightforward. A naive approach transfers full-precision weights and activations to processing units and quantizes each tensor twice, resulting in much more memory movement than expected.
Additional overhead comes from the activation tensors, whose sizes grow substantially because of the \imcol\ transformation applied before quantization. 
We propose \MicroQonv, a way to combine microscaling with convolutional layers' forward and backward operations by quantizing each tensor only once and quantizing the activation tensor before applying a modified version of \imcol: channel-batch-first \imcol. \MicroQonv\ reduces the quantization cost by a factor of $\times2$ for weights and gradients, and by up to $\times9$ for activations, at a negligible accuracy cost. 
It reduces memory movement and storage by up to $\times7.53$ compared to their full-precision counterparts.
This way, MicroQonv reduces microscaling-quantized activation memory movement by $\times3.5$ for state-of-the-art object detection models YOLOV8nano and $\times2.2$ for YOLOV26nano. It also enables 4-bit microscaling in a quantized latent replay strategy for continual learning at the edge, improving accuracy by +5.7\% to +11\%. 

\end{abstract}


\section{Introduction}



Recent progress in Artificial Intelligence (AI) has enabled substantial improvements in numerous fields, such as natural language processing or medical imaging. Some of its applications are executed on resource-constrained devices, e.g., UAVs \cite{cereda_-device_2024} and autonomous driving systems, while models are mostly trained in the cloud. Whether the model is trained in the cloud or on an edge device, backpropagation is energy-intensive and time-consuming: reducing training costs will benefit all training scenarios. Moreover, giving embedded systems the ability to train locally is a huge improvement to their autonomy. Thanks to Continual Learning (CL)~\cite{wang_comprehensive_2023}, models trained in the cloud can be adapted to the dynamic real-world environment encountered by each device, and can even learn new tasks and classes while maintaining performance on previously seen data. A simple yet very effective CL strategy is the replay buffer~\cite{rolnick2019experience}, where devices regularly store data from their environment to keep training on it while encountering new data.

By reducing the number of bits per parameter for weights, activations, and gradients, quantization techniques~\cite{nagel_white_2021} help reduce energy and latency cost in two complementary ways: they reduce memory movement~\cite{gholami_ai_2024} to bring data to the processing elements, and lower processing cost by using smaller hardware operators~\cite{benaliDATE24,leran2025mpicc}. 
Thus, combining training strategies with quantization — also known as Fully-Quantized Training (FQT) — is a good approach to enhancing device efficiency.

Recent microscaling ~\cite{rouhani_shared_2023, rouhani_microscaling_2023} data formats stand out among quantization methods because they support 4-bit quantization, often using the 4-bit Floating-Point (FP4) format to maintain near-full-precision accuracy. For example, Large Language Models (LLMs) are trained from scratch using various flavors of FP4 and microscaling for weights, activations, and gradients~\cite{chmiel_fp4_2025, nvidia_pretraining_2025}. 
Microscaling partitions a tensor into line-shaped blocks whose elements share a single scaling factor. For efficiency, these lines must be aligned with the direction of the dot-product computation. However, weights, activations, and gradients participate in two distinct General Matrix Multiplications (GeMMs), and their computation directions differ. As a result, they must be quantized twice, which is suboptimal.
The issue is particularly pressing for activations: the tensor is expanded via explicit or implicit \imcol\ (image-to-column)~\cite{dukhan_indirect_2019,zhou_characterizing_2021} before being quantized to the microscale format. Since the same full-precision element can be assigned different quantized values due to varying distributions inside the blocks, 
either the \imcol-expanded quantized tensor or the original full-precision tensor must be saved for back-propagation, which results in excessive memory movement and memory storage compared to other quantization techniques. This contributes to the well-known "AI memory wall" \cite{gholami_ai_2024} problem.

This paper addresses these issues of increased memory movement and quantization computation due to the microscaling format by proposing the following contributions. We first present \MicroQonv, a tensor reordering method that quantizes each tensor only once during training, using the microscaling format. Activations are quantized before an implicit variant of \imcol\ called \textit{channel-batch-first \imcol} is applied, significantly reducing memory movement, memory storage, and overhead due to quantization computations.
\MicroQonv\ is compared with other quantized training methods, showing its relevance by achieving higher accuracy at the same quantization level or by requiring much less computation with only a tiny accuracy loss, far outweighed by the memory size and memory movement reductions it provides. The gains are validated using a DRAM simulator \cite{dramsim3}, which simulates memory transfers between off-chip and on-chip memory. Notably, MicroQonv reduces microscaling-quantized activation memory movement in state-of-the-art object detection models, by $\times3.5$ for YOLOV8nano and $\times2.2$ for YOLOV26nano.
We then apply 4-bit \MicroQonv\ to a Continual Learning setting using quantized latent replay buffer strategy, to improve the state-of-the-art (SoTA) final model accuracy by up to +11\%.

The paper is organized as follows. Section~\ref{sec:background} gives the background on training, \imcol\ transformations, 
and quantization. In Section~\ref{sec:motivations}, we describe the problems that arise when applying microscaling quantization to convolutional layers. Section~\ref{sec:microqonv} describes the \MicroQonv\ approach. Then, Section \ref{sec:dramsimulation} presents the DRAM simulation results to quantify energy and latency gains. Section~\ref{sec:results} presents the experimental setup and results in a deep learning scenario, while Section~\ref{sec:continuallearning} shows improvements in a continual learning scenario.



\section{Background and Related Work} \label{sec:background}




\subsection{CNN Training}

Following the emergence of the Transformer architecture for natural language processing, Vision Transformers (ViTs)~\cite{dosovitskiy2021image} gained widespread popularity in computer vision. Because ViT achieves very good accuracy with the attention mechanism, it is worth asking whether convolutional layers will remain state-of-the-art. As an example, ConvNext~\cite{Woo2023ConvNeXtVC} demonstrates that carefully designed convolutional networks can compete with vision transformers in accuracy. More importantly, embedded systems applications such as image classification and object detection not only require accuracy but also low latency and low energy consumption. In that domain, the high computational cost of attention mechanisms can be a disadvantage, as shown in~\cite[Fig. 1]{mobileone}. In object detection, as we write these lines, YOLO26~\cite{jocher2026ultralyticsyolo26unifiedrealtime} and RF-DETR~\cite{robinson2026rfdetr} achieve the best accuracy-latency tradeoff depending on the deployment platform. 
Like earlier YOLO versions, YOLO26 is a fully convolutional network. While RF-DETR mainly uses a transformer architecture, it also uses a YOLO convolutional projector to have richer encoder features. Another competitive model is RT-DETRv4~\cite{liao2025rtdetrv4painlesslyfurtheringrealtime}, which combines a convolutional backbone with an attention mechanism. Regarding real-time open vocabulary object detection, convolution-based YOLO26-E~\cite{jocher2026ultralyticsyolo26unifiedrealtime} is leading the way. Thus, in embedded-system applications, the convolutional layer is a core component, and improving it will benefit the overall system.

The training process of convolutional neural networks (CNNs) involves three primary operations performed on each convolution layer: \textbf{forward propagation}, \textbf{backward propagation}, and \textbf{weight update}. In this subsection, we focus on standard convolution layers. For clarity, we assume a stride of~1 and no padding. However, the methods presented in this paper are independent of these assumptions.

\subsubsection{{Forward Pass}}
Given a tensor representing a batch of $N$ input feature maps $X \in \mathbb{R}^{N \cdot C_{in}  \cdot H_{in}  \cdot W_{in}}$,\footnote{''$\cdot$`` denotes starting a new dimension, e.g. here $X$ is a 4D tensor.} and convolutional weights $W \in \mathbb{R}^{C_{out} \cdot C_{in} \cdot K_h \cdot K_w}$, where $C_{in}$ and $C_{out}$ are the input and output channel dimensions, respectively, $H_{in} \cdot W_{in}$ the input dimensions, and $K_h \cdot K_w$ the convolution kernel dimensions, the output feature map tensor is computed using the convolution operation:
\begin{equation}
    Y = W \star X,
    \label{eq:forward}
\end{equation}
where $\star$ denotes the standard convolution. 
The output tensor is defined as
$
    Y \in \mathbb{R}^{N \cdot C_{out} \cdot H_{out} \cdot W_{out}},
$
with
$
    H_{out} = H_{in} - K_h + 1$ and
$ 
    W_{out} = W_{in} - K_w + 1.
$


\subsubsection{{Backward Pass}}
During backpropagation, compute two gradients: the gradient with respect to the input activations and the gradient with respect to the weights.

\noindent \textit{Input Gradient:}
Given the gradient of the output feature maps
$
    G_Y \in \mathbb{R}^{N \cdot C_{out} \cdot H_{out} \cdot W_{out}},
$
the gradient with respect to the input activations is computed by convolving $G_Y$ with the transposed kernels:
\begin{equation}
    G_X = W^{T} \star G_Y,
    \label{eq:gx}
\end{equation}
where
$
    W^{T} \in \mathbb{R}^{C_{in} \cdot C_{out} \cdot K_h \cdot K_w}
$
is obtained by swapping the input and output channel dimensions of $W$.

\noindent \textit{Weight Gradient:}
The gradient w.r.t the weights
$
    G_W \in \mathbb{R}^{C_{out} \cdot C_{in} \cdot K_h \cdot K_w}
$
is computed by accumulating over the batch dimension:
\begin{equation}
    G_W[c_{out}, c_{in}, :, :] =
    \sum_{n=1}^{N}
    \left(
        G_Y[n, c_{out}, :, :] \star
        X[n, c_{in}, :, :]
    \right),
    \label{eq:gw}
\end{equation}
with $c_{in} \in [1\dots C_{in}]$ and $c_{out} \in [1\dots C_{out}]$. Notations in this paper follow the principles of Python tensor iterators.

\subsubsection{{Weight Update}}
After computing the gradients, the weights are updated using a learning rate~$\eta$. A standard Stochastic Gradient Descent update step is computed as
$$    W = W - \eta \, G_W.  $$

\subsection{Compute Architecture Model} \label{sec:comparch}
\begin{figure}[htbp]
    \includegraphics[width=1.0\columnwidth]{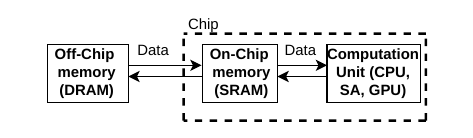}
    \caption{Generic compute architecture model considered.}
    \label{fig:comparch}
\end{figure}

CNN training computation can run on different hardware targets, such as a Central Processing Unit (CPU), a Graphics Processing Unit (GPU), or a Systolic Array (SA). While these architectures differ in performance and use cases, they share one limitation: none can store the entire model in the memory of the chip performing the computations. Thus, they require frequent data movement to and from off-chip memory to load GeMM inputs and store GeMM outputs. The general and hardware-agnostic architecture considered in this paper is illustrated in Figure ~\ref{fig:comparch}. Computations are performed on data (weights, activations, or gradients) in the on-chip memory. Data must be transferred to and from off-chip memory before computation. We consider optimal reuse of tiles and tensors once located in on-chip memory.

\subsection{Im2col Transformation} \label{sec:im2col}
The \imcol\ transformation is the most widely used technique for mapping convolution operations to general matrix multiplications~\cite{dukhan_indirect_2019,zhou_characterizing_2021}. This enables the use of highly optimized matrix-multiply libraries or dedicated hardware units (e.g., systolic arrays). In this subsection, we illustrate the forward and backward pass transformations for a standard convolution layer, a necessary step for understanding \microqonv\ as detailed in Section \ref{sec:microqonv}.
Given the input activation tensor $X$, the goal of the \imcol\ operation is to rearrange all local receptive fields (i.e., patches) of size $C_{\text{in}} K_h K_w$ from all $N$ input feature maps into the columns of a large 2D matrix $X' = \imcol(X)$.

\subsubsection{{Input Rearrangement}}
For each sample $n \in \{1,\ldots,N\}$ and for every spatial output position $(h, w)$, corresponding to $h \in \{1,\dots,H_{out}\}$ and $w \in \{1,\dots,W_{out}\}$, the \imcol\ operator extracts the patch of $X[n, :, h : h + K_h - 1,\; w : w + K_w - 1]$.  
This patch is then \emph{flattened} into a vector of length $C_{in} K_h K_w$ and placed as one column of the resulting matrix $X' \in \mathbb{R}^{(C_{in} K_h K_w)\cdot(N H_{out} W_{out})}$.

\noindent Therefore, \textbf{Rows} of $X'$ correspond to all kernel positions across all input channels.  
\textbf{Columns} of $X'$ correspond to every spatial position of the sliding window across all samples in the batch.

\subsubsection{{Kernel Flattening}}
The convolution kernel $W$
is similarly reshaped into a 2D matrix $ W_{f} \in \mathbb{R}^{C_{out} \cdot (C_{in} K_h K_w)}$ by flattening each filter $W[c_{out}, :, :, :]$ into a row vector of length $C_{in} K_h K_w$, preserving the channel-major layout:
$$
    W_{f}[c_{out}, :] 
    = \mathrm{flatten}\!\left(W[c_{out}, :, :, :]\right).
$$

\subsubsection{{Matrix Multiplication Formulation}}
After applying both transformations, the forward convolution in Equation \ref{eq:forward} can be computed as a single matrix multiplication $Y_f = W_{f} \cdot X'$,
 where
 $
 Y_{f} \in \mathbb{R}^{C_{out} \cdot (NH_{out}W_{out})}
 $
 is a flattened representation of the output tensor $Y$.

The \imcol\ and flattened formats are used in the backward pass for computing the flattened weight gradients $G_{W_{f}} \in \mathbb{R}^{C_{out} \cdot (C_{in} K_h K_w)}$: 
\[
G_{W_{f}} = G_{Y_{f}} \cdot X'^T
\]
where $G_{Y_{f}} \in \mathbb{R}^{C_{out} \cdot (N H_{out} W_{out})}$ is a flattened 2D representation of the output gradient $G_Y$ and $X'^T$ the transpose matrix of $X'$.
Consequently, the input gradient $G_{X'}$ with the same format as $X'$
is also computed using matrix multiplication:
\[
G_{X'} = W_{f}^T \cdot G_{Y_{f}}.
\]

This classic \imcol\ approach is central to high-performance CNN implementations, enabling convolutions to reuse efficient GeMM kernels and hardware acceleration. However, it incurs significant memory overhead, as the reduced feature matrix from the \imcol\ transformation can use up to $K_h \times K_w$ times as much memory as the original feature map.

\subsubsection{{Channel-First Im2col}}
\begin{figure}[htbp]
    \includegraphics[width=1.0\columnwidth]{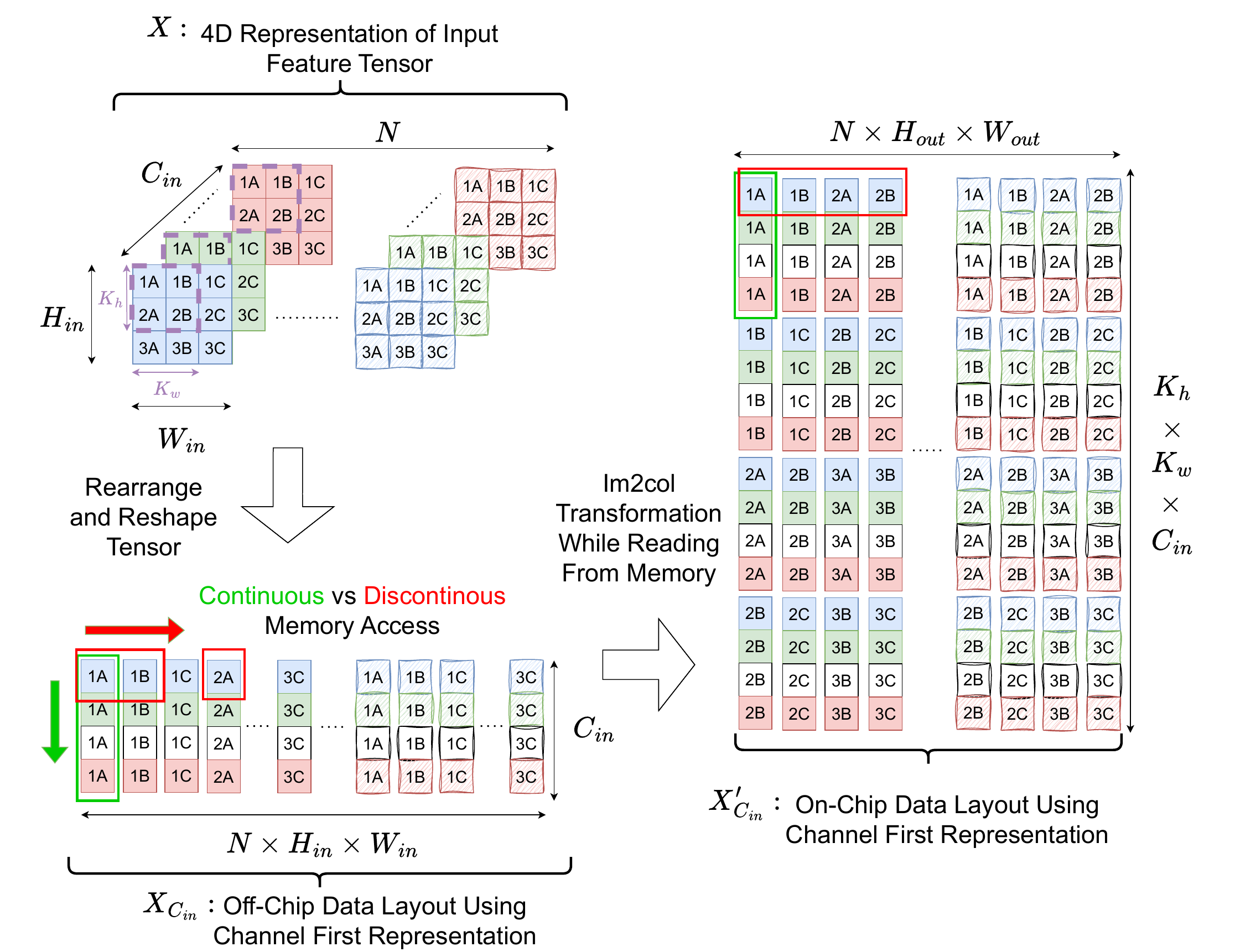}
    \caption{Example of the channel-first layout for a convolution layer with $K_h = K_w = 2$ and $H_{in}=H_{out}=3$. The tensor is flattened along the spatial and batch dimensions, producing $N H_{in} W_{in}$ contiguous representation across columns, each containing all $C_{in}$ channels. These columns are streamed to generate the \imcol\ matrix. Forming the transposed representation requires reading across lines rather than columns.}
    \label{fig:channel_first}
\end{figure}

In \cite{zhou_characterizing_2021}, the authors introduce a more efficient variant of the \imcol\ transformation that relies on storing the input feature maps in a channel-first layout within off-chip memory. As illustrated in Figure~\ref{fig:channel_first}, the input tensor is organized as
$
    X_{C_{in}} \in 
    \mathbb{R}^{C_{in} \cdot (N H_{in} W_{in})},
$
rather than being stored directly in the full \imcol\ format $X'_{C_{in}}$.
This data layout enables the hardware to contiguously load input activations into on-chip memory \emph{along the input channel dimension} as the convolution window slides over the spatial domain. Consequently, the \imcol\ columns can be generated \emph{on the fly}, during the read operation. By avoiding explicit \imcol\ storage and exploiting contiguous channel-wise access, this approach significantly reduces both off-chip memory usage and bandwidth requirements.

However, this optimization applies only to the \emph{forward} convolution pass. During backpropagation, the computation of the weight gradients requires loading the \emph{transposed} activation matrix $X_{C_{in}}'^T$, which inherently incurs discontinuous access patterns across the $W_{in}$ dimension of the channel-first tensor $X_{C_{in}}$ stored in off-chip memory. This discontinuous access pattern is illustrated by the red arrow in Figure~\ref{fig:channel_first}. As a consequence, the channel-first layout does not preserve the same efficiency for the backward pass, limiting this approach to inference or forward-only execution paths. 

Other \imcol\ improvement methods exist, such as~\cite{fornt2023ontheflyim2col, cho2017memoryefficientconvolution}: they generate \imcol\ elements as the computation unit needs them, avoiding storage of the \imcol-expanded tensor. While this is efficient for FP32 tensors where elements are independent, these methods do not combine well with microscaling quantization. Because the same activation pixel repeated in \imcol\ shares several microscaling blocks, each time with different neighboring pixels, it can have several different microscaling-quantized values in the \imcol\ tensor. Thus, the same element must be re-quantized several times, as in classic \imcol\ expansion. Furthermore, either the FP32 tensor must be stored in memory for reuse in backward propagation, or its microscaling-quantized \imcol\-expanded version must be stored, which significantly reduces the off-chip-to-on-chip memory-movement gains of microscaling quantization.

To address these limitations, Section \ref{sec:microqonv} introduces a new storage layout designed to maintain contiguous memory access in both the forward and backward passes of convolution layers, enabling efficient end-to-end training without relying on explicit \imcol\ representations within off-chip memory.\footnote{{As shown in Figure~\ref{fig:batch_first}, we keep the following notation for the rest of the paper: when a dimension of a 2D matrix is defined by a product of tensor dimensions, the last dimension of the product is the one along which memory access is performed.}}

\subsection{Quantization} \label{sec:quant}

Quantization strategies aim to reduce the quantization error between full-precision elements and their quantized versions. The goal is to balance accurate quantization of small values (i.e., rounding error) with accurate quantization of large values (i.e., clipping error). A small rounding error causes a large clipping error, and vice versa.
%
This is where microscaling~\cite{rouhani_microscaling_2023, rouhani_shared_2023} thrives. By performing per-block quantization that does not clip the block's maximum value, microscaling improves the balance between rounding and clipping errors. In blocks with large values, the clipping error is minimized at the expense of the approximation of small elements of that block. In other blocks, rounding error is minimized. 
This balance between clipping and rounding errors makes microscaling a good technique for quantizing weights, activations, and gradients~\cite{chmiel_fp4_2025}, enabling energy-efficient low-precision multiply-and-accumulate (MAC) units.

In the original microscaling format~\cite{rouhani_microscaling_2023}, the block scale is a power of two encoded in E8M0 format. The NVFP4 format from Nvidia \cite{nvidia_pretraining_2025} further reduces quantization error by performing per-tensor scaling before the block-scale calculation in the 8-bit floating-point FP8 E4M3 (4-bit exponent, 3-bit mantissa plus one sign bit) format. 
The added complexity of per-tensor quantization and the mantissa in the block scale is offset by lower quantization error, enabling small-bitwidth formats and the hardware gains they provide while maintaining satisfactory accuracy.
Elements are quantized in FP4 E2M1 as in the original microscaling format~\cite{rouhani_microscaling_2023}, organized as block-lines of $B=16$ elements.
In the experiments, we assume a master copy of the weights is kept in FP32 precision and used to update the weights, as in SoTA methods.

\section{Motivations} \label{sec:motivations}
We first describe how to quantize a tensor in the microscaling format, then show its limitations in convolution operations, which prior work has not addressed.

\subsection{{Microscaling quantization}}
To quantize following the microscaling format, a tensor $X$ is decomposed into blocks  $X_i$ of $B$ elements $\{x_{i,1},\dots,x_{i,B}\}$.  
Following the principles of the NVFP4 format, each value of $X$ is scaled with a per-tensor scaling factor $T_{scale}$ and a per-block scaling factor $B_{scale}$.
\begin{equation}\label{eq:rescale_tensor}
    T_{scale} = \max |X| / \left(\max B_{sf} \times \max B_{ef}\right),
\end{equation}
where $\max B_{sf}$ is the maximum of the block-scale format (i.e., 448 for FP8 E4M3), and $\max B_{ef}$ is the maximum of the element format (i.e., 6 for FP4 E2M1). 
\begin{equation} \label{eq:block_max}
B_{scale} = CastE4M3\left(\max |X_i|/(T_{scale}  \times max B_{ef}) \right),
\end{equation}
where $CastE4M3(.)$ rounds its argument to the nearest FP8 E4M3 value. Block elements $X_i$ are turned into quantized format $X_i^q$ as
\begin{equation} \label{eq:block_quantize}
X_i^q = Round\left(X_i /(B_{scale} \times T_{scale})\right),
\end{equation}
where $Round(.)$ denotes rounding, either round-to-nearest or stochastic, to an FP4 E2M1 format value. In simulation, fake quantization is obtained by multiplying $X_i^q$ by $B_{scale} \times T_{scale}$.
Following~\cite{chmiel_accurate_2021}, activations and weights are rounded to the nearest quantized value while stochastic rounding is performed on the gradient. 

\subsection{{Issues with dot products of microscaled blocks}}
We consider two vectors $R=\{r_1,\dots,r_S\}$ and $T=\{t_1,\dots,t_S\}$ of size $S$, which belong to bigger matrices quantized using microscaling with block size $B$. If the quantization is performed in the same direction as the dot product, the dot product can be written as 
\begin{equation} \label{eq:goodcomputationblock}
Q(R\times T) = Q\left(\sum_{i=1}^S r_{i} \times t_{i}\right) = \sum_{j=1}^{\lceil S/B \rceil} sr_j \times st_j \times \sum_{i=1}^B r^{q}_{i,j} \times t^{q}_{i,j}
\end{equation}
where $\lceil . \rceil$ rounds the content to the nearest superior integer, and $Q(.)$ represents quantization of both vectors involved in microscaling format. $sr_{j}$ and $st_{j}$ represent the scales of each microscale block, \(r^q_{i,j}\) and \(t^q_{i,j}\) are quantized elements of $R$ and $T$ vectors, respectively, in the $i$-th position of the $j$-th block. 

However, when the quantization direction is not the one of the dot product, scale blocks must be applied after every low-precision element multiplication, 
making the computation inefficient to the point where requantizing the whole tensor in the same direction is better.
Ideally, each tensor would be quantized only once. In practice, as seen in the matrix multiplication formulation of Section~\ref{sec:im2col} for $G_{X'}$, $Y_f$ and $G_{W_f}$, each tensor is involved in two GeMM computations that happen in different directions: weight and activation matrices are transposed in the backward pass, whereas output gradients are on different sides of the GeMM product.

\subsection{{Limitations of related work}}
The associated code with the DaCapo accelerator~\cite{kim_dacapo_2024} shows that their convolutional layers 
load full precision tensors and then quantize them in line, in the forward as well as in the backward pass. To avoid requantization and extra memory movement, the accelerator of Cuyckens \textit{et al.}~\cite{cuyckens_efficient_2025} uses square quantization: instead of quantizing along a single dimension, it quantizes in a square shape across two dimensions of the tensor. Thus, tensors quantized in the forward pass can be reused in the backward pass. 
However, the authors consider only models with fully-connected layers. Extrapolating their method to convolutional layers involves memorizing the square-quantized version of an \imcol-expanded tensor for activations during the forward pass, for its reuse during the backward pass. This reduces, if not cancels, the memory-movement and memory storage gains from quantization. Therefore, combining microscaling quantization and convolution layer operations remains an open question.

\section{\MicroQonv: Reshaping Memory Layout for Efficient Microscaling Training} 
\label{sec:microqonv}
In this section, we first introduce channel-batch-first \imcol, a method that reorders weights, activations, and gradients so they can be quantized once during training while maintaining efficient computation. We then specify the selected quantization format, show the quantization and memory-movement gains, and compare \microqonv's accuracy with other methods.

\subsection{{Reshaping memory layout with channel-batch-first im2col}}
\begin{figure}[b]
    \centering
    \vspace{-0.3cm}\hspace{-0.1cm}
    \includegraphics[width=1.1\columnwidth]{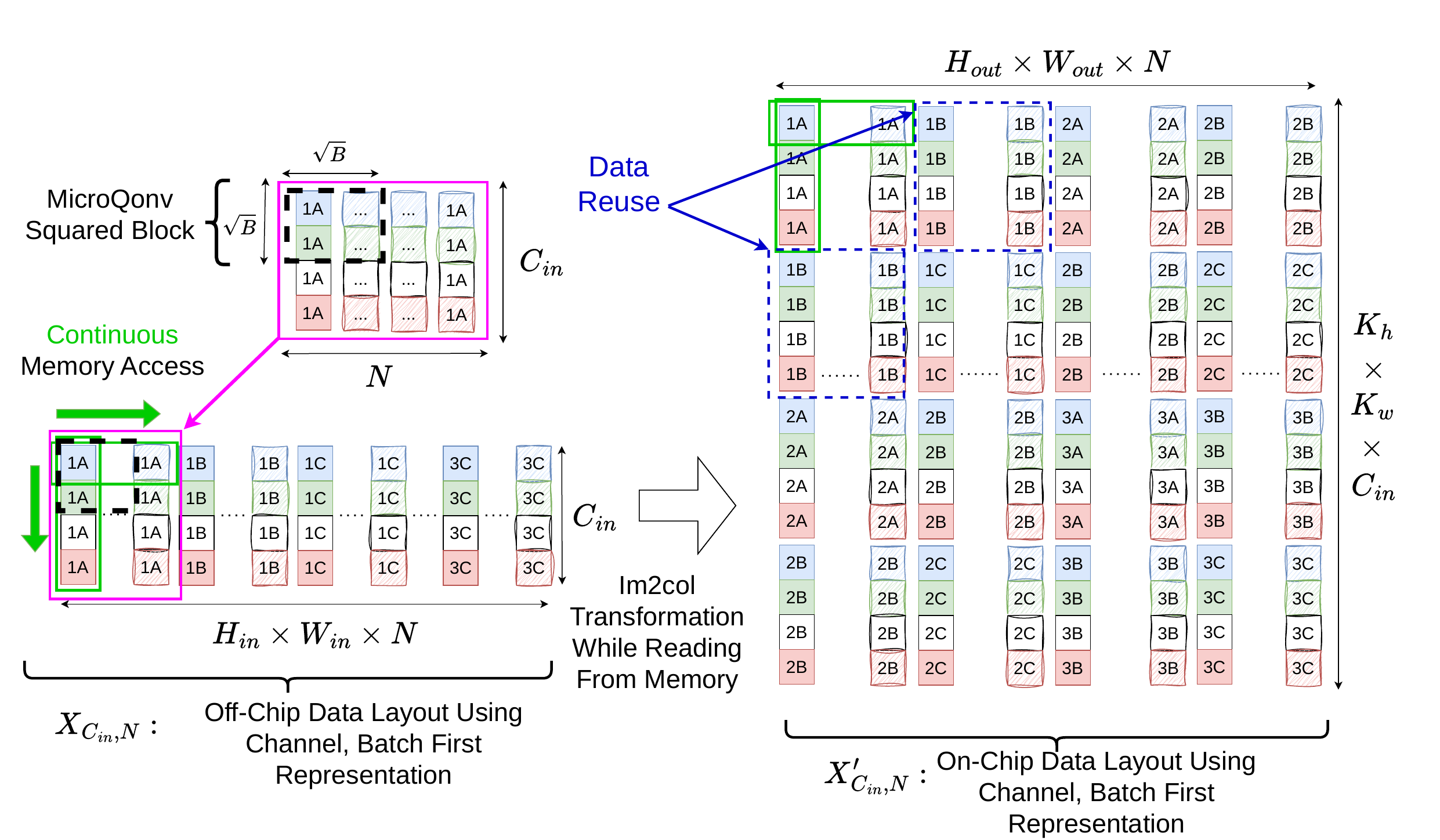}

    \caption{Channel-batch-first layout: activations are stored compactly in off-chip memory for contiguous \imcol\ and transpose access. Highlighted squares show MicroQconv quantization blocks across $N$ and $C_{in}$ dimensions. This layout also enables activation reuse, reducing redundant memory reads by avoiding repeated accesses to the same data.}
    \label{fig:batch_first}
\end{figure}

We store the input feature maps in a layout that enables \emph{contiguous data access} along both the input-channel dimension $C_{in}$ and the batch dimension $N$. To this end, we propose rearranging the input activations $X$, as presented in Figure \ref{fig:channel_first}, to $X_{C_{in},N} \in \mathbb{R}^{C_{in} \cdot (H_{in} W_{in} N)}$, as illustrated in Figure \ref{fig:batch_first}. In this representation, each spatial location $(h, w)$ contains a contiguous block of all channels across all batch elements.
This organization ensures contiguous memory access during both major phases of convolutional training. During the forward pass, streaming the \imcol\ matrix $X'_{C_{in},N} \in \mathbb{R}^{(K_h K_w C_{in}) \cdot (H_{out} W_{out} N)}$ into on-chip memory requires sliding the convolution window and reading consecutive values along the $C_{in}$ dimension. Since channel values are stored contiguously at each spatial position, the \imcol\ blocks can be generated on the fly with minimal memory overhead. During the backward pass, computing input gradients requires reading the transposed activation matrix $X^T_{C_{in},N}$ along the batch dimension $N$. With batch elements stored adjacently, these accesses are also contiguous, eliminating the discontinuous striding that occurs in channel-first layouts. 

This layout preserves the memory-storage benefits of the channel-first representation while enabling optimal memory-bandwidth use for both forward and backward convolution operations. Furthermore, this layout enables data reuse: contiguous blocks of $X_{C_{in},N}$ in either channel or batch dimension can be loaded only once from off-chip memory to on-chip memory and repeated on-chip to form channel-batch first \imcol\ representation $X'_{C_{in},N}$, noticeably reducing memory movement.

\subsection{{Efficient quantization directions}}
Reordering input activations with channel-batch-first \imcol\ requires changing the shapes of other tensors for GeMM computations. $W_f$ must be reshaped as $W_{f C_{out},C_{in}} \in \mathbb{R}^{C_{out} \cdot (K_{h} K_{w}C_{in})}$, and $G_{Y_{f}}$ reshaped as $G_{Y_{f} C_{out},N} \in \mathbb{R}^{C_{out} \cdot (H_{out}W_{out}N)}$. We now have a 2D matrix formulation of all tensors, which are compliant with all three GeMM computations: $Y_{f C_{out},N}=W_{f C_{out},C_{in}}\cdot X'_{C_{in},N}$, $G_{W_{f} C_{out},C_{in}}=G_{Y_{f} C_{out},N}\cdot X'^T_{C_{in},N}$ and $G_{X' C_{in},N}=W^T_{f C_{out},C_{in}}\cdot G_{Y_{f} C_{out},N}$. Simple reordering of $Y_{f C_{out},N}$, $G_{W_{f} C_{out},C_{in}}$ and $G_{X' C_{in},N}$ turns them into $Y_{f}$, $G_{W_{f}}$ and $G_{X'}$ respectively. Note that $X'_{C_{in},N}$ is the \imcol\ transformation of off-chip data layout $X_{C_{in},N}$, while $W_{f C_{in},C_{out}}$ and $G_{Y_{f} C_{out},N}$ have the same representation off-chip and on-chip. 

With a 2D matrix formulation, we can quantize each tensor into a square block, so the computation direction always aligns with one dimension of the square. This ensures contiguous memory accesses and efficient computations as in Eq.~\ref{eq:goodcomputationblock} with a small variation: block scales must be applied every $\sqrt{B}$ elements in both dimensions covered by the square. This way, $X_{C_{in},N}$ is quantized along $C_{in}$ and $N$ directions, $W_{f C_{in},C_{out}}$ is quantized along $C_{in}$ and $C_{out}$ directions and $G_{Y_{f} C_{out},N}$ is quantized along $C_{out}$ and $N$ directions. 
Figure~\ref{fig:batch_first} shows an example of \MicroQonv\ quantization block for input activations in 2D shape, as stored in the memory: quantizing $X_{C_{in},N}$ allows for keeping contiguous memory access for quantization blocks after \imcol\ transformation into $X'_{C_{in},N}$. 

We combine these quantization directions with the NVFP4 format described in Section \ref{sec:quant}, except the block shape is a square, not a line, and its size is $B=64$ $(8\times8)$ as in~\cite{cuyckens_efficient_2025}, not 16. All experiments involving microscaling quantization are with a block size of $B=64$. 
While we employ a specific block size and representation, \MicroQonv\ is compliant with all block scales and element representation formats, and applies to all microscaling variants, including their early version, the block floating-point format.

\MicroQonv\ is hardware-agnostic: it applies to every hardware accelerator following the generic model depicted in Figure~\ref{fig:comparch}. Among others, it applies easily to custom fully-quantized-training accelerators with an explicit off-chip/on-chip hierarchy and programmable address generation/dataflow, the same class as Dacapo~\cite{kim_dacapo_2024} and Cuyckens \textit{et al.}~\cite{cuyckens_efficient_2025}. \MicroQonv\ is a design technique informing how such accelerators should lay out and quantize data, and it requires zero additional hardware circuitry. 


\subsection{{Gains in memory movement and per-element quantization}}
Table~\ref{tab:quantization_memmove_cost} compares \MicroQonv\ performance with Dacapo~\cite{kim_dacapo_2024}, and an adaptation of Cuyckens \textit{et al.}~\cite{cuyckens_efficient_2025} to convolutional layers, whose methods are available in Section~\ref{sec:motivations}. The method described in Cuyckens \textit{et al.} quantizes the \imcol-expanded activation tensor in square 8$\times$8 blocks and saves it for reuse during backward propagation. Dacapo quantizes the \imcol-expanded tensor in line blocks of 64, following a constant line direction for forward propagation and a constant column direction for backward propagation, and stores the FP32 activation tensor. These theoretical results are obtained by applying a standard convolution using layer dimensions with parameters: $K_{H}=K_{W}=3$, padding and stride of 1. Per-element (Per-elt.) quantization represents the number of times an element is quantized. Backward tensor size represents the size of the tensor that must be transferred and stored off-chip during forward propagation, to then be brought on-chip during backward propagation.
The backward tensor size is compared to the \MicroQonv\ tensor.
As discussed previously, \MicroQonv\ reordering allows for quantizing every tensor once, and the reordered activation tensor $X_{C_{in},N}$ is quantized before its \imcol\ expansion into $X'_{C_{in},N}$. Table~\ref{tab:quantization_memmove_cost} confirms the claims with reductions of the activation quantization overhead by $\times9$ and the backward memory movement for activations by $\times7.53$ compared to other methods, which would translate into significant energy gains for the hardware performing training. This is further analyzed in Section \ref{sec:dramsimulation}.
\begin{table}[h!]
\centering
\caption{Quantization overhead and memory movement during backpropagation for Weights (W), Activations (A) and Gradients (G) of a convolutional layer with $K_h=K_w=3$ quantized with 4-bit microscaling. }
    
\label{tab:quantization_memmove_cost}
\begin{tabular}{c|ccc|cc} 
\hline
 \textbf{Method}  &\multicolumn{3}{|c|}{\makecell{\textbf{Per elt.}\\ \textbf{quantization}}}& \multicolumn{2}{|c}{\makecell{\textbf{Backward}\\ \textbf{Mem. Move.}}} \\
 & A & W & G & A & W \\
\hline
    $\text{Dacapo}$  \cite{kim_dacapo_2024}  & $\times18 $     & $\times2$   & $\times2$  &  $\times 7.53 $ &  $\times7.53 $   \\
    Cuyckens \textit{et al.} \cite{cuyckens_efficient_2025}  &    $\times9$     & $\times1 $     & $\times1 $    & $\times9$  & $\times1$    \\
    $\text{\textbf{\MicroQonv}}$   &   $\times\textbf{1}$& $\times\textbf{1} $& $\times\textbf{1} $ & $\times\textbf{1} $     & $\times\textbf{1}$  \\
\hline
\end{tabular}
\end{table}


    




\subsection{{Accuracy analysis}}
All methods use NVFP4 quantization format and a block size of 64, unless otherwise specified. Since \MicroQonv\ introduces a new quantization pattern, the computation differs slightly from other methods. 
To compare accuracy, we train a ResNet-32 model on the CIFAR-100 dataset and a ResNet-18 model on the first 100 ImageNet classes. All methods use the same hyperparameters. Both training methods use a cosine-annealed scheduler and stochastic gradient descent with a momentum of 0.9. ResNet-32 is trained with a learning rate of 0.1 for 200 epochs, a weight decay of $5e^{-4}$, and a batch size of 128. ResNet-18 is trained with a learning rate of 0.05 for 150 epochs, a weight decay of $4e^{-5}$, and a batch size of 256. All layers are quantized except the first and last layers, which are kept at full precision.
Results reported in Table~\ref{tab:accuracy_mse_error_quantization} include the test accuracy mean and standard deviation ($\sigma$) on five training runs. All methods are within 0.6\% of the FP32 baseline in both training scenarios; the different quantization methods are within a 0.25\% $\sigma$ window, which is roughly the margin of error. \MicroQonv\ is slightly below Cuyckens \textit{et al.}, itself slightly below Dacapo~\cite{kim_dacapo_2024}.

\begin{table}[b]
\centering
\caption{Comparison of test accuracy and quantization MSE per element on ResNet-32 (R32) with CIFAR-100 (C100) and Resnet-18 (R18) with ImageNet (Inet100).}
\label{tab:accuracy_mse_error_quantization}
\setlength{\tabcolsep}{3pt} 
\begin{tabular}{c|cc|cccc} 
\hline
 \textbf{Method} & \multicolumn{2}{c|}{\textbf{Test acc \%}} & \multicolumn{3}{c}{\textbf{Normalized MSE}} \\
   & R32-C100 & R18-Inet100 & A & W & G  \\
\hline
    FP32 baseline   &  $71.05\pm.29$& $75.55\pm.22$ & $NA$ & $NA$ & $NA$  \\
    Dacapo \cite{kim_dacapo_2024}     &  $70.86\pm.24$& $75.19\pm.12$  & $0.891$ & $0.982$ & $0.827$  \\
    \makecell{Cuyck. \textit{et al.} \cite{cuyckens_efficient_2025}} &   $70.82\pm.11$  & $75.20\pm.18$ &$0.875$  & $0.992$ & $0.996$     \\
    \textbf{\MicroQonv}        &  $70.62\pm.13$&  $75.08\pm.36$ & $1.0$  & $1.0$ & $1.0$  \\
    \textbf{\microqonvp}        &  $71.10\pm.28$& $75.76\pm.12$  & $0.236$  & $1.0$ & $1.0$  \\
\hline
\end{tabular}
\end{table}

We further investigate this by measuring the average per-element quantization mean-square error (MSE) for Weights (W), Activations (A), and output Gradients (G), and normalize values by \MicroQonv\ error for readability, on Resnet-32 with CIFAR-100.
While all methods have similar weight errors, the activation MSE is slightly higher for \MicroQonv. Since the same element is repeated several times in the \imcol-expanded tensor, it is quantized several times, inducing a slightly lower average quantization error than when quantized once, as in \MicroQonv.
It is interesting to note that in-line (1D) gradient quantization error~\cite{kim_dacapo_2024} is lower than its square (2D) counterpart of \cite{cuyckens_efficient_2025} and \MicroQonv, which is probably why in-line quantization has moderately better accuracy than square quantization. 
However, this tiny accuracy loss must be balanced against the energy gains provided by \microqonv\ shown in Section~\ref{sec:dramsimulation}.
In energy-limited scenarios, spending less energy means training for more epochs, yielding better accuracy within the same energy budget. In these cases, it is likely that \MicroQonv\ will close the accuracy gap with the other methods thanks to significant energy savings from reduced memory movement and quantization overhead. 
Furthermore, if accuracy is extremely important, we propose \microqonvp, in which activations are quantized in NVFP6 (identical to the NVFP4 format, except that elements are quantized in FP6 E3M2). As shown in Table~\ref{tab:training_accuracy}, while weights and gradients remain in NVFP4, its accuracy exceeds that of the other methods and matches the FP32 baseline. Although \microqonvp\ requires $\approx$50\% more memory movement for activations because it uses 6 bits per element instead of 4 bits for \MicroQonv, it still requires less memory movement and storage: roughly $\times$5.1 less than Dacapo and $\times$6 less than Cuyckens \textit{et al.} extrapolation, while providing better accuracy, as shown in Table~\ref{tab:accuracy_mse_error_quantization}. 


\subsection{{Viability in Object Detection}}
We train YOLOV8nano and YOLOV26nano using Ultralytics\footnote{https://github.com/ultralytics/ultralytics/releases/tag/v8.4.14} setup and native hyperparameters, except batch size, which is set to 8. Both models are pretrained on the COCO dataset and finetuned for 20 epochs on the PascalVOC dataset for YOLOV8nano~\cite{pascal-voc-2012}, while YOLOV26nano is finetuned on the KITTI dataset~\cite{Geiger2013kitti} for 20 epochs. To maintain a mean Average Precision (mAP50 and mAP50-95 \cite{mAPref2016}) close to the baseline, all methods use the NVFP8 representation for weights, activations, and gradients, which is the same as NVFP4 except that elements are in FP8 E4M3 format instead of FP4 E2M1. 
For YOLOV8nano, we keep the first layer and the last Distribution Focal Loss (DFL) convolution layer~\cite{xiangli2020generalizedfocalloss} at full precision. For YOLOV26nano, only the first layer is not quantized. YOLOV26nano also includes depthwise convolution layers, which are more sensitive to quantization and account for less than 1\% of YOLOV26nano's MACs. Therefore, we keep them in BF16 precision.
The mean and standard deviation for mAP50 and mAP50-95 across five runs are shown in Table~\ref{tab:object_detection_accuracy} for YOLOV8nano and Table~\ref{tab:8bit_objdet_acc_y26kitti} for YOLOV26nano. We append '-8' to the method names to indicate the use of the NVFP8 format. 

For YOLOV8nano, \MicroQonv-8 is within 0.3 mAP of the FP32 baseline and less than 0.15 away from the Dacapo-8 and Cuyckens et al.-8 extrapolations, while offering much better energy efficiency, as we show in the following section. We further report in Table~\ref{tab:object_detection_accuracy} an experiment in which the model is trained from scratch for 100 epochs on \MicroQonv\ and the FP32 baseline only. Here, \MicroQonv-8 is within 0.2 mAP of the baseline.

Regarding YOLOV26nano, microscaling quantization slightly improves the FP32 baseline by providing regularization, which prevents overfitting. \MicroQonv-8 is above Dacapo-8 and Cuyckens et al.-8. We also train \MicroQonv\ from scratch for 100epochs and compare it to the FP32 baseline. There, \MicroQonv-8 is .2 mAP away from the baseline with the mAP50 metric and less than 0.8mAP away from FP32 accuracy with the mAP50-95 metric.

For both YOLO models, the close distance between \MicroQonv-8 accuracy and FP32 accuracy indicates that \MicroQonv\ can enhance the efficiency of SoTA object detection models.


\begin{table}[htbp]
\centering
\caption{Object detection performance on YOLOV8nano model and PascalVOC dataset.}
\label{tab:object_detection_accuracy}
\setlength{\tabcolsep}{2pt} 
\begin{tabular}{c@{ }|c@{ }c|c@{ }c} 
\hline
 \textbf{Method} & \multicolumn{2}{c|}{\textbf{Finetuning}} & \multicolumn{2}{c}{\textbf{From scratch}}  \\
   & mAP50  & mAP50-95 & mAP50  & mAP50-95 \\
\hline
    FP32 baseline   &  $82.06\pm.06$& $60.89\pm.06$&  $76.53\pm.12$   &  $55.56\pm.10$   \\
    {\MicroQonv-8}        &  $81.87\pm.15$&  $60.62\pm.12$   &  $76.52\pm.27$&  $55.39\pm.19$ \\
    Dacapo-8 \cite{kim_dacapo_2024}     &  $81.88\pm.10$ &  $60.75\pm.10$& - & - \\
    \makecell{Cuyck. \textit{et al.}-8} &   $81.85\pm.06$ &   $60.75\pm.11$  & -  & -  \\
\hline
\end{tabular}
\end{table}

\begin{table}[htbp]
\centering
\caption{Object detection performance on YOLOV26nano model and KITTI dataset.}
\label{tab:8bit_objdet_acc_y26kitti}
\vspace{-0.2cm}
\setlength{\tabcolsep}{2pt} 
\begin{tabular}{c@{ }|c@{ }c|c@{ }c} 
\hline
 \textbf{Method} & \multicolumn{2}{c|}{\textbf{Finetuning}} & \multicolumn{2}{c}{\textbf{From scratch}}  \\
   & mAP50  & mAP50-95 & mAP50  & mAP50-95 \\
\hline
    FP32 baseline   &  $53.91\pm.40$& $33.75\pm.24$&  $74.26\pm1.2$   &  $49.44\pm.73$   \\
    {\MicroQonv-8}        &  $55.10\pm1.3$&  $34.20\pm.78$   &  $74.11\pm1.2$&  $48.74\pm.55$ \\
    Dacapo-8 \cite{kim_dacapo_2024}     &  $54.44\pm.47$ &  $33.56\pm.89$& - & - \\
    \makecell{Cuyck. \textit{et al.}-8} &  $54.88\pm1.3$ &  $33.90\pm.79$  & -  & -  \\
\hline
\end{tabular}
\end{table}

\section{DRAM Simulation}\label{sec:dramsimulation}

\begin{figure}
\includegraphics[width=0.8\columnwidth]{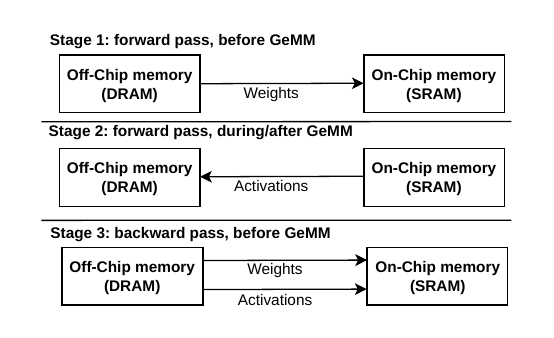}
    \caption{Memory transfers considered for DRAM simulation.}
    \label{fig:memtransfer}
\end{figure}

To measure latency reduction and energy gains from reduced memory movement, we simulate memory movement using DRAMsim3~\cite{dramsim3}. 
Cuyckens \textit{et al.} store the entire model on-chip, which is realistic for their use cases where models consist of a few fully-connected layers, but not for much larger convolutional networks. We assume that we can store input and output tensors for one layer on-chip, while other layer tensors must be stored off-chip. We consider memory movement during forward and backward pass phases, as shown in Figure~\ref{fig:memtransfer}. In the forward pass, the activation tensor is computed as the output of the last layer. Therefore, it is already on-chip. Because we quantize weights during the weight-update phase to reduce memory movement, we only need to transfer the quantized weights (if a quantization method is employed, as in Cuyckens \textit{et al.} extrapolation and \MicroQonv, but not in Dacapo). We then quantize the activations to perform the forward convolution, and save the quantized activations in memory. During backward propagation, the gradient tensor is already on-chip due to the previous layer's computation. The system then loads weights and activations on-chip for the associated backward operations.

We exclusively consider transfers between off-chip (DRAM) and on-chip (SRAM) memories, as memory movement between off-chip and on-chip memory is much more costly than internal on-chip memory movement and computations, as shown in~\cite{horowitz2014energyproblem}. Because all methods use the same convolution operations, internal on-chip memory transfer and computation costs will not differ much between methods, unlike off-chip-to-on-chip transfers.

\subsection{Experimental Setup}

We simulate various model and dataset configurations: ResNet32 with CIFAR-100, ResNet18 with ImageNet, YOLOV8nano with PascalVOC, and YOLOV26nano with KITTI. 
Batch size is 32 for CIFAR100 (image size of $32\times32\times3$) and ImageNet ($224\times224\times3$), and 8 for PascalVOC and KITTI ($640\times480\times3$). All reported results are obtained using the LPDDR4 (Low-Power Double Data Rate) memory model from the DRAMsim3 repository\footnote{https://github.com/umd-memsys/DRAMsim3/releases/tag/1.0.0}.


\subsection{Results}
Table~\ref{tab:memory_energy_gain} reports the tensor size transferred between off-chip memory and on-chip memory, the transfer energy spent from DRAMsim3, and the memory transfer latency from DRAMsim3, for the whole model. As predicted by the analytical model in Table~\ref{tab:quantization_memmove_cost}, \MicroQonv\ shows improvements over both methods.
Given the same quantization method and the same model, the gains of these three parameters are the same. We have observed this not only with LPDDR4 but also with DDR5, DDR6, and HBM (High Bandwidth Memory) memory models. For simplicity, while we refer to energy in the following discussion, the same statements apply to tensor size and memory transfer latency.

The gain of \MicroQonv\ compared to saving full-precision activations like in Dacapo is more or less constant and close to $\times7.53$ as in Table \ref{tab:quantization_memmove_cost} for ResNet-32 and ResNet-18, close to $\times3.93$ for YOLOV8n and close to $\times3.4$ for YOLOV26n. Since the activation and weight tensors are of the same dimensions in both cases, the gain comes from the quantization format. DRAM overhead varies slightly across methods because of their internal components, which explains the small variation around that gain. The improvement factor is closer to $\times3.93$ for YOLOV8n because elements are quantized in FP8 instead of FP4. Improvement factor is around $\times3.4$ for YOLOV26n because of depthwise convolution layers which are quantized in BF16 for both methods.
    
For Cuyckens \textit{et al.} extrapolation, however, the \imcol-expanded version is saved, and its dimensions heavily depend on the stride and kernel sizes of the tensor. Roughly, the expansion factor of \imcol\ is $K_h\times K_w$/stride. Thus, a smaller stride and a larger kernel size will increase the gains of \MicroQonv\ over \imcol\ expansion, and vice versa. In these conditions, all ResNet32 layers have kernel sizes $K_h=K_w=3$. Most layers have a stride of 1, and a few have a stride of 2, which explains the energy gain of around $\times$8.5 in activation size, energy, and latency. ResNet18 has 17 convolutional layers with a kernel size of 3$\times$3, 14 with a stride of 1, and 3 with a stride of 2, which explains why the energy gain is closer to $\times7$. 
In the YOLOv8n model, most layers have a stride of 1; 40\% of them use a kernel size of 1$\times$1, and it regroups most layers with large activation tensors. Other layers use a 3$\times$3 kernel, which explains the energy gain of about $\times3.5$. In the YOLOv26n model, most layers use a stride of 1 too. The percentage of layers using a kernel size of 1 increases to 53\%, and a few layers with a kernel size of 3 are depthwise convolution layers, which are quantized in BF16. That is why YOLOv26n has an energy gain of about $\times2.2$.
Regarding the weights, as there is no \imcol\ expansion, Cuyckens \textit{et al.} and \MicroQonv\ thus quantize the same tensor. Only the quantization pattern varies, which explains the lack of difference.

 
This simulation does not account for quantization cost: when the kernel size is larger than 1, the \imcol-expanded tensor will require quantizing roughly $K_h\times K_w$/stride more elements, which in turn requires more silicon area for quantization or adds extra latency, further improving the gains of \MicroQonv.



    

\begin{table}[htbp]
\centering
\caption{Energy (E) in mJ and Latency (L) in ms for all transfers in Fig.~\ref{fig:memtransfer}. Tensor Size (S) is in MegaBytes (MB). }
\label{tab:memory_energy_gain}
\setlength{\tabcolsep}{3pt} 
\begin{tabular}{c|ccc|ccc} 
\hline
 \textbf{Method}&\multicolumn{3}{c}{\textbf{Activations}}&\multicolumn{3}{c}{\textbf{Weights}}\\
  R32-Cifar100 & S(MB) & E(mJ) & L(ms) & S(MB) & E(mJ) & L(ms)         \\
\hline
    Dacapo                              &38.67& 92.56  & 46.63 &  1.87 & 4.402 &1.95 \\
    \makecell{Cuyck. \textit{et al.}} &42.57& 100.40 & 50.89  &  0.244 & 0.529&0.164 \\
    \textbf{\MicroQonv}                  &5.04& 11.68  & 6.09   &  0.244 & 0.529&0.164 \\

\hline
  R18-Imagenet & S(MB) & E(mJ) & L(ms) & S(MB) & E(mJ)  & L(ms)          \\
\hline
    Dacapo                              &279.44& 669.68 & 337.91 & 46.72 & 111.9& 56.37\\
    \makecell{Cuyck. \textit{et al.}} &271.92 & 643.91 & 324.97 &  6.28 &14.77 &7.33 \\
    \textbf{\MicroQonv}                  &38.56 & 91.12  & 45.97  & 6.28  &14.77 & 7.33\\
\hline
  YV8n-PVOC & S(MB) & E(mJ)  &  L(ms)  & S(MB) & E(mJ) &  L(ms)       \\
\hline

    Dacapo-8                              &492.44 & 1179.85  & 595.19  & 12.58   & 29.9  &14.76 \\
    \makecell{Cuyck \textit{et al.}}-8    &441.13 & 1057.15  & 533.46  & 3.195   & 7.44  &3.35 \\
    {\MicroQonv-8}                  &125.22 & 300.03   & 151.36  & 3.195   & 7.44  &3.35 \\

\hline
  YV26n-KITTI & S(MB) & E(mJ)  &  L(ms)  & S(MB) & E(mJ) &  L(ms)       \\
\hline

    Dacapo-8                              & 634.82 & 1520.73  & 767.43  & 10.17   & 23.98 & 11.56 \\
    \makecell{Cuyck \textit{et al.}}-8    & 420.39 & 1007.26  & 507.96 & 2.61 & 5.95  & 2.45\\
    {\MicroQonv-8}                  & 187.16 & 448.33  & 225.85 & 2.61 & 5.95 & 2.45 \\

\hline

\end{tabular}
\end{table}

\section{Training Performance}\label{sec:results}

We first demonstrate the relevance of \MicroQonv\ by comparing its performance with other fully-quantized training techniques in a classic learning scenario, which yields more stable results to facilitate comparison. Experiments are conducted on image classification, then on object detection. In all tables, quantization methods use the same setup, including learning rate and number of epochs, for a fair comparison.

\subsection{Image classification}

We train ResNet-32 and ResNet-18 using the same setup as in Table~\ref{tab:accuracy_mse_error_quantization}. 
Table~\ref{tab:training_accuracy} compares final test accuracy (mean and standard deviation over five training runs) between the FP32 baseline, LUQ \cite{chmiel_accurate_2021}, QLR \cite{ravaglia_tinyml_2021}, and \MicroQonv. 4-bit per-tensor and per-channel quantization \cite{nagel_white_2021} showed more than 10\% accuracy degradation or convergence issue, they are therefore not displayed.
In LUQ~\cite{chmiel_accurate_2021}, activations, weights, and gradients are quantized with a per-tensor scale. 
In ResNet-32 and ResNet-18, all convolutions except the first one occur after a ReLU activation function: the sign bit can be replaced by a mantissa/exponent bit for activations, or simply removed, leaving activations at 3-bit precision. We report LUQ results in two cases: keeping the original paper configuration at 4 bits (UINT4) or matching the \MicroQonv\ effective bitwidth at 3 bits (INT4, LUQ$\dagger$). In the first case, \MicroQonv\ shows better average performance with only half as many quantization bins. In the second case, \MicroQonv\ beats LUQ by more than 2\% for ResNet-32 and by 0.8\% for ResNet-18. \MicroQonv\ also uses a unified format, leading to better accelerator efficiency per silicon area than LUQ, which requires different computation formats for forward and backward propagation.
Hence, \MicroQonv\ delivers the best precision at 4-bit quantization. Its FP4 E2M1 activation format saves 3 bits per activation pixel instead of 4 (+8/64 = 3.125 with block scale), significantly increasing the amount of data we can store.
In Table~\ref{tab:training_accuracy}, we also compare to QLR \cite{ravaglia_tinyml_2021}, where training is done in FP32 for weights, activations, and gradients, then activations and weights are quantized to 8 bits for inference. QLR provides slightly better accuracy, at a much higher training cost than \MicroQonv. 
\begin{table}[h!]
\centering
\setlength{\tabcolsep}{3pt} 
\caption{Performance comparison of quantization methods on ResNet-32 (R32) with CIFAR-100 (C100) and Resnet-18 (R18) with ImageNet (Inet100). FP4E corresponds to the E3M0 format.}
\label{tab:training_accuracy}
\begin{tabular}{c|ccc|cc} 
\hline
 \textbf{Technique} & \multicolumn{3}{c|}{\textbf{Elements Format}} & \multicolumn{1}{c}{\textbf{Test acc \%}} \\
  & \textbf{A} & \textbf{W}  & \textbf{G}  & R32-C100 & R18-Inet100 \\
\hline
    $\text{Baseline}$                 &  $\text{FP32}$      & $\text{FP32}$       & $\text{FP32}$       &  $71.05\pm.29$  & $75.55\pm.22$  \\
\hline

    Dacapo \cite{kim_dacapo_2024}     & FP4 & FP4 & FP4 & $70.86\pm.24$& $75.19\pm.12$   \\
    \makecell{Cuyck. \textit{et al.} \cite{cuyckens_efficient_2025}} & FP4 & FP4 & FP4 &   $70.82\pm.11$  & $75.20\pm.18$ \\
    $\text{\MicroQonv}$                &  $\text{FP4}$  & $\text{FP4}$   & $\text{FP4}$   &  $70.62\pm.13$ & $75.08\pm.36$  \\
    $\text{\microqonvp}$                &  $\text{FP6}$  & $\text{FP4}$   & $\text{FP4}$   &  $71.10\pm.28$  & $75.76\pm.12$  \\
\hline
    $\text{QLR}$ \cite{ravaglia_tinyml_2021} &  $\text{UINT8}$     & $\text{INT8}$       & $\text{FP32}$       &  $70.95\pm.41$  & $75.28\pm.25$  \\
    $\text{LUQ}$  \cite{chmiel_accurate_2021}                    &  $\text{UINT4}$     & $\text{INT4}$       & $\text{FP4E}$   &  $70.54\pm.35$ & $74.77\pm.22$   \\
    $\text{LUQ$\dagger$}$                      &  $\text{INT4}$      & $\text{INT4}$       & $\text{FP4E}$   &   $68.32\pm.08$ & $74.19\pm.10$    \\

\hline
\end{tabular}
\end{table}

\subsection{Object detection}

YOLOV8nano and YOLOV26nano are trained reusing Table~\ref{tab:object_detection_accuracy} and Table~\ref{tab:8bit_objdet_acc_y26kitti} setup. In addition to MicroQonv, Dacapo, and Cuyckens et al., we evaluate accuracy using symmetric per-tensor and per-channel quantization. Per-tensor quantization uses one scale per tensor, mapping the tensor's absolute maximum value to the absolute maximum of the quantized element format for weights, activations, and gradients. Per-channel quantization attributes one scale per output channel for weight and gradient tensors, and one scale per input channel for activation tensors. The absolute maximum value of a channel is mapped to the quantized element absolute maximum.
All methods use FP6 E2M3 for element quantization. We append '-6' to the method names to indicate the use of the FP6 format. This means microscaling-based quantization methods use NVFP6 representation, identical to NVFP4 except that elements are in FP6 E2M3 format instead of FP4 E2M1. 

Because the input and output layers are more sensitive to quantization, we keep the input layer and every output layer in higher precision. Because the YOLOV8 model aggregates 3 levels of feature maps and has two branches for prediction, one for localization and one for prediction, it has 6 output layers plus the DFL layer kept in full precision (out of 64 convolution layers in the model). YOLOV26 further divides each branch into one-to-one (used for training only) and one-to-many (used for training and inference) subbranches, in order to get rid of Non-Maximum-Suppression (NMS) step \cite{jocher2026ultralyticsyolo26unifiedrealtime}: its 12 output layers (out of 126 convolution layers in the model) remain in full precision. Following Table~\ref{tab:8bit_objdet_acc_y26kitti} setup The depthwise convolution layers of YOLOv26nano are set to BF16 precision. Results are shown in Table~\ref{tab:6bit_objdet_acc_y8pvoc} and Table~\ref{tab:6bit_objdet_acc_y26kitti}. 

For YOLOV8nano, \MicroQonv-6 is within 0.3mAP of the baseline and less than .15mAP away from Dacapo-6 and Cuyckens et al.-6. More importantly, it is significantly above Per-tensor-6 and 0.4mAP to 0.8mAP higher than Per-channel-6. 
\begin{figure}[htbp]
    \centering
    \begin{subfigure}[b]{1\columnwidth}
        \includegraphics[width=\columnwidth]{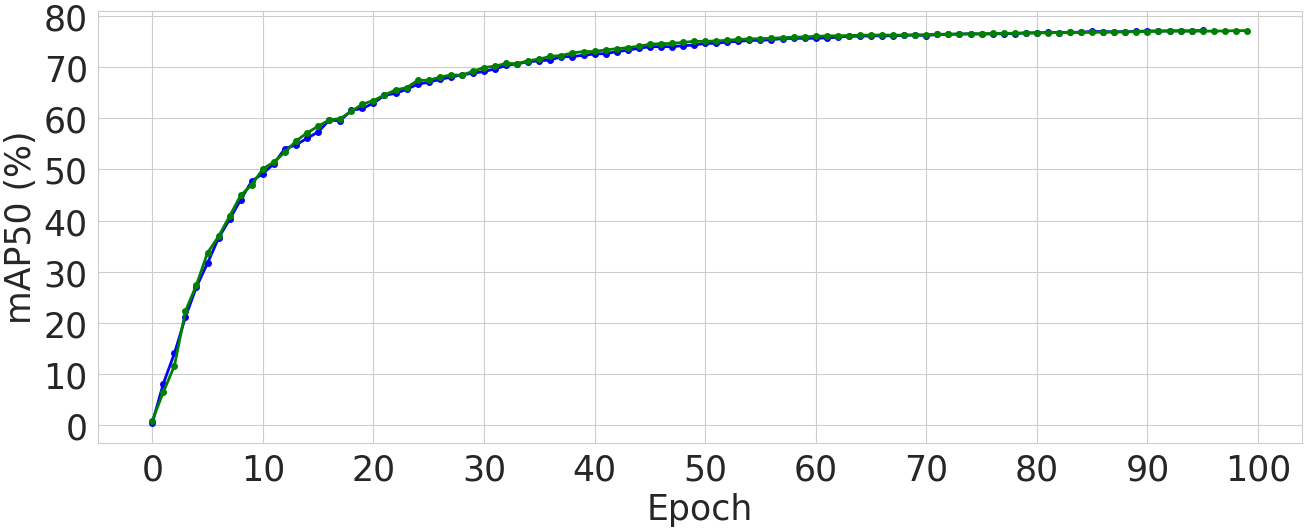}
        \caption{Test of accuracy during training on YOLOV8nano using the PascalVOC dataset.}
        \label{fig:train_curve:top}
    \end{subfigure}
    
    \begin{subfigure}[b]{1\columnwidth}
        \includegraphics[width=\columnwidth]{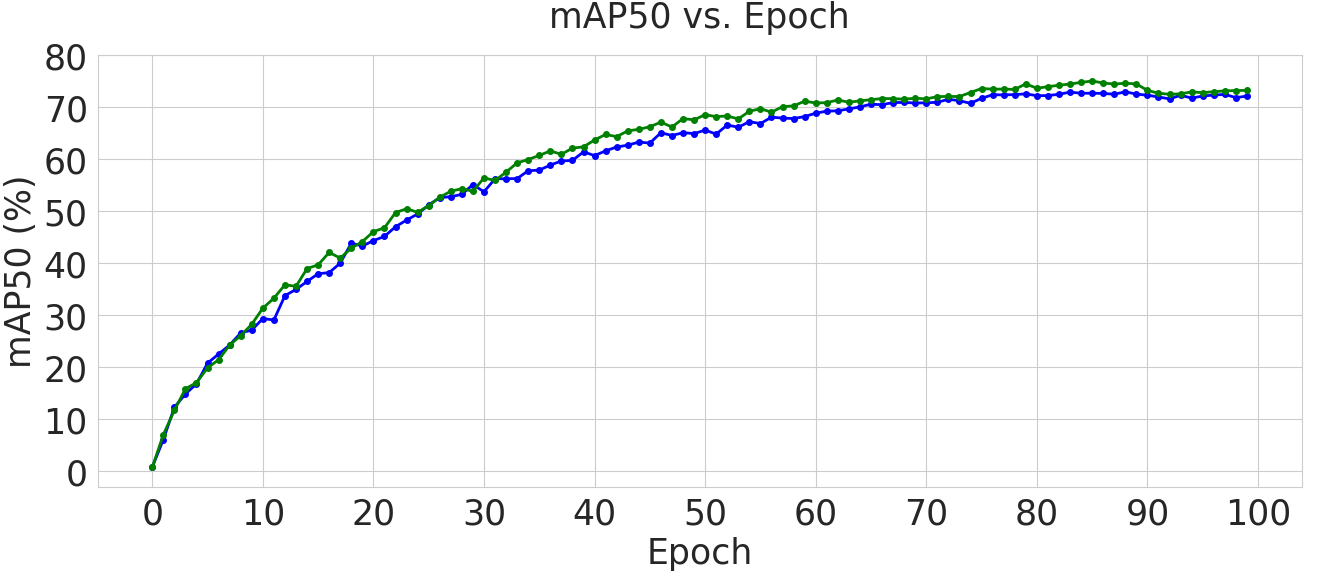}
        \caption{Test of accuracy during training on YOLOV26nano using the KITTI dataset.}
        \label{fig:train_curve:bottom}
    \end{subfigure}
    \caption{Test of accuracy along training of \MicroQonv-6 (in blue) and FP32 baseline (in green).}
    \label{fig:train_curve}
\end{figure}
A similar pattern emerges from the YOLOV26nano results: \MicroQonv-6 is less than 0.3mAP from the FP32 baseline in the fine-tuning scenario and less than 1.3mAP away in the training-from-scratch scenario. It shows a 0.1mAP improvement over Dacapo-6 and is less than 0.85mAP away from Cucykens et al.-6, which is above the FP32 baseline. At the same time, it is way higher than per-tensor quantization and 1.2mAP to 3mAP higher than per-channel quantization.

Figure~\ref {fig:train_curve} shows the test-of-accuracy training-curve comparison between \MicroQonv-6 and the FP32 baseline in a training-from-scratch scenario for YOLOV8nano using the PascalVOC dataset (Fig.~\ref {fig:train_curve:top}) and YOLOV26nano using the KITTI dataset (Fig.~\ref {fig:train_curve:bottom}).
As shown in the figure, \MicroQonv-6 closely follows the FP32 curve across epochs.

Section~\ref{sec:dramsimulation} has shown that \MicroQonv is significantly more energy efficient than Dacapo and Cuyckens et al., and this section shows that it also maintains accuracy close to these methods and to the FP32 baseline. Moreover, \MicroQonv is more accurate and offers a much more hardware-friendly computation pattern than per-channel quantization~\cite{nagel_white_2021}. It therefore has a better accuracy-energy efficiency tradeoff than every other quantization method.


\begin{table}[htbp]
\centering
\caption{6 bit Object detection performance on YOLOV8nano-PascalVOC.}
\label{tab:6bit_objdet_acc_y8pvoc}
\setlength{\tabcolsep}{2pt} 
\begin{tabular}{c@{ }|c@{ }c|c@{ }c} 
\hline
 \textbf{Method} & \multicolumn{2}{c|}{\textbf{Finetuning}} & \multicolumn{2}{c}{\textbf{From scratch}}  \\
   & mAP50  & mAP50-95 & mAP50  & mAP50-95 \\
\hline
    FP32 baseline   &  $82.33\pm.02$& $61.33\pm.09$&  $77.11\pm.29$   &  $56.25\pm.22$   \\
    {\MicroQonv-6}        &  $82.12\pm.20$&  $61.08\pm.13$   &  $76.95\pm.26$&  $55.98\pm.26$ \\
    Per Channel-6        &  $81.31\pm.14$&  $60.15\pm.12$   &  $76.58\pm.20$&  $55.55\pm.15$ \\
    Per Tensor-6        &  $65.88\pm.42$&  $43.89\pm.35$   &  - &  - \\
    Dacapo-6 \cite{kim_dacapo_2024}     &  $82.19\pm.10$ &  $61.14\pm.06$& - & - \\
    \makecell{Cuyck. \textit{et al.}-6} &  $82.18\pm.09$ &  $61.20\pm.13$  & -  & -  \\
\hline
\end{tabular}
\end{table}

\begin{table}[htbp]
\centering
\caption{6 bit Object detection performance on YOLOV26nano-KITTI.}
\label{tab:6bit_objdet_acc_y26kitti}
\setlength{\tabcolsep}{2pt} 
\begin{tabular}{c@{ }|c@{ }c|c@{ }c} 
\hline
 \textbf{Method} & \multicolumn{2}{c|}{\textbf{Finetuning}} & \multicolumn{2}{c}{\textbf{From scratch}}  \\
   & mAP50  & mAP50-95 & mAP50  & mAP50-95 \\
\hline
    FP32 baseline   &  $53.91\pm.73$& $33.86\pm.61$&  $74.26\pm1.2$   &  $49.44\pm.73$   \\
    {\MicroQonv-6}        &  $54.11\pm.93$&  $33.51\pm.51$   &  $73.02\pm.62$&  $48.18\pm.54$ \\
    Per Channel-6        &  $51.22\pm.64$&  $31.49\pm.39$   &  $71.96\pm1.1$&  $46.99\pm.81$ \\
    Per Tensor-6        &  $9.1\pm.76$&  $4.09\pm.42$   &  - &  - \\
    Dacapo-6 \cite{kim_dacapo_2024}     &  $54.08\pm.11$ &  $33.36\pm.63$& - & - \\
    \makecell{Cuyck. \textit{et al.}-6} &  $54.93\pm.65$ &  $34.02\pm.46$  & -  & -  \\
\hline
\end{tabular}
\end{table}

\section{\microqonv\ in a Continual Learning Setting}\label{sec:continuallearning}


Continual learning (CL)~\cite{wang_comprehensive_2023} aims to train AI models continually to adapt to changing data distributions. Because the complete history of data seen by the model is too large to be stored entirely, especially for on-device models, and because backpropagation optimizes only on currently available data, catastrophic forgetting occurs. Thus, the field aims to mitigate catastrophic forgetting. 
We focus on the Class Incremental Learning (CIL) scenario \cite{vandeVen2022}, where a model is trained sequentially on a series of tasks, each introducing new classes not seen in previous tasks. 
In this section, we present results in a (CL) setting that relies on latent replay~\cite{pellegrini_latent_2020}.



\subsection{Related Work}
Among existing CL strategies, the replay buffer \cite{rolnick2019experience} is one of the simplest and most effective ways to mitigate catastrophic forgetting. At the end of each task, a subset of the training data is selected and stored in the replay buffer memory. The replay buffer is then combined with data from subsequent tasks, allowing the model to learn new classes while mitigating catastrophic forgetting of previously seen classes. It can be used alone or in conjunction with other strategies to ensure optimal performance~\cite{zhou_memo_2022, kwon_lifelearner_2023}. 

\begin{figure}[h]
    \centering
    \includegraphics[width=0.8\columnwidth]{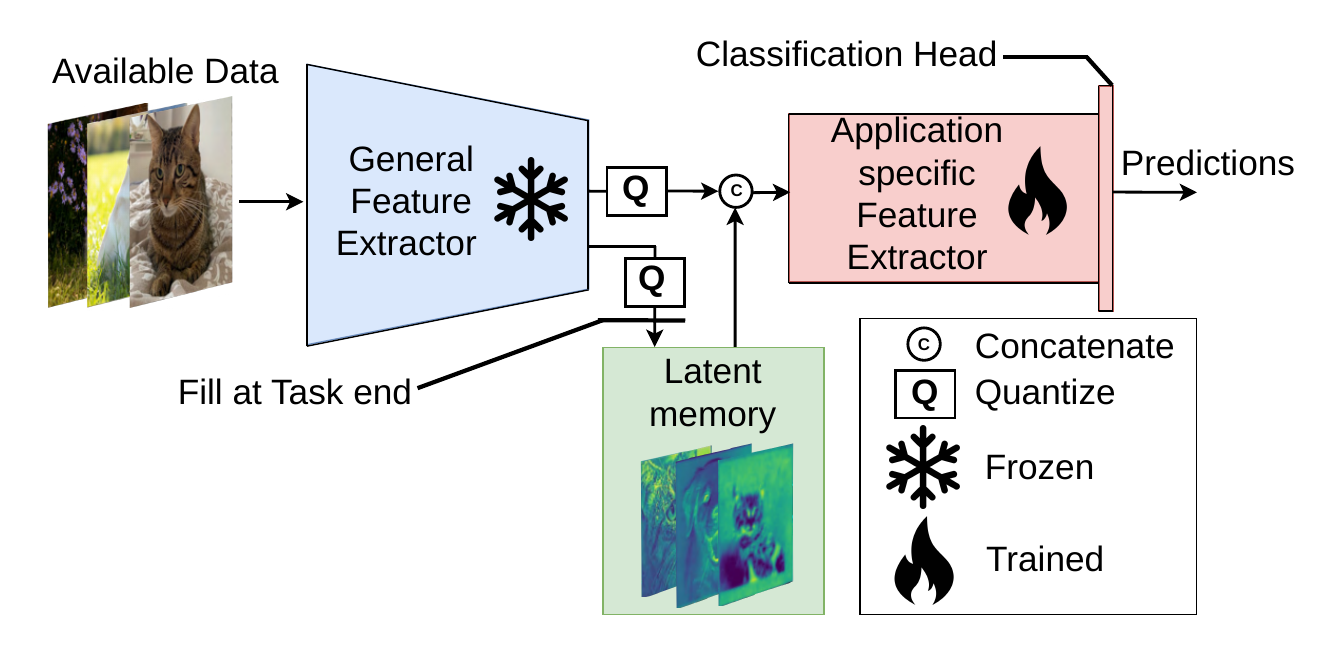}
    \caption{Quantized Latent Replay (QLR) strategy overview. Feature extractor layers are also quantized.}
    \label{fig:latent-replay}
\end{figure}

The latent replay buffer~\cite{pellegrini_latent_2020} (LR), as shown in Figure~\ref{fig:latent-replay}, is a variant of the replay buffer: early layers of the feature extractor are frozen, and the output activations of the frozen feature extractor, rather than the original images, are stored in the replay buffer. This significantly reduces backpropagation computational cost. Moreover, early layers tend to learn generic features from large datasets~\cite{miro-panades_772jframe_2024}, limiting accuracy degradation compared to whole-model retraining. 
Compressing the replay buffer items allows for storing more elements for the same memory budget, without LR~\cite{wang_memory_2022,yang_probing_2024,luo_class-incremental_2023} or with LR~\cite{hayes_remind_2019}. However, full-precision models used in these papers make these methods hard to adapt to energy-constrained devices.

Few papers study the interaction between quantization and the latent replay buffer. Input activations from a convolutional or linear layer can be quantized and stored in memory to further reduce memory storage. LifeLearner~\cite{kwon_lifelearner_2023} combines 8-bit quantization with a meta-learned feature extractor and product quantization. However, they require pre-training on the same data distribution as the test data, making them sensitive to distribution shifts. Quantized Latent Replay (QLR)~\cite{ravaglia_tinyml_2021} combines LR with integer 8-bit quantization for weights and activations during inference, while training is done in FP32. They use no compression technique other than quantization, which gives them good adaptability to unseen distributions. We seek to improve these results by using the \textit{ad hoc} microscaling format proposed by \MicroQonv. 

\subsection{Experimental Results}
\begin{figure}[bp]
    \centering
    \includegraphics[width=0.95\columnwidth]{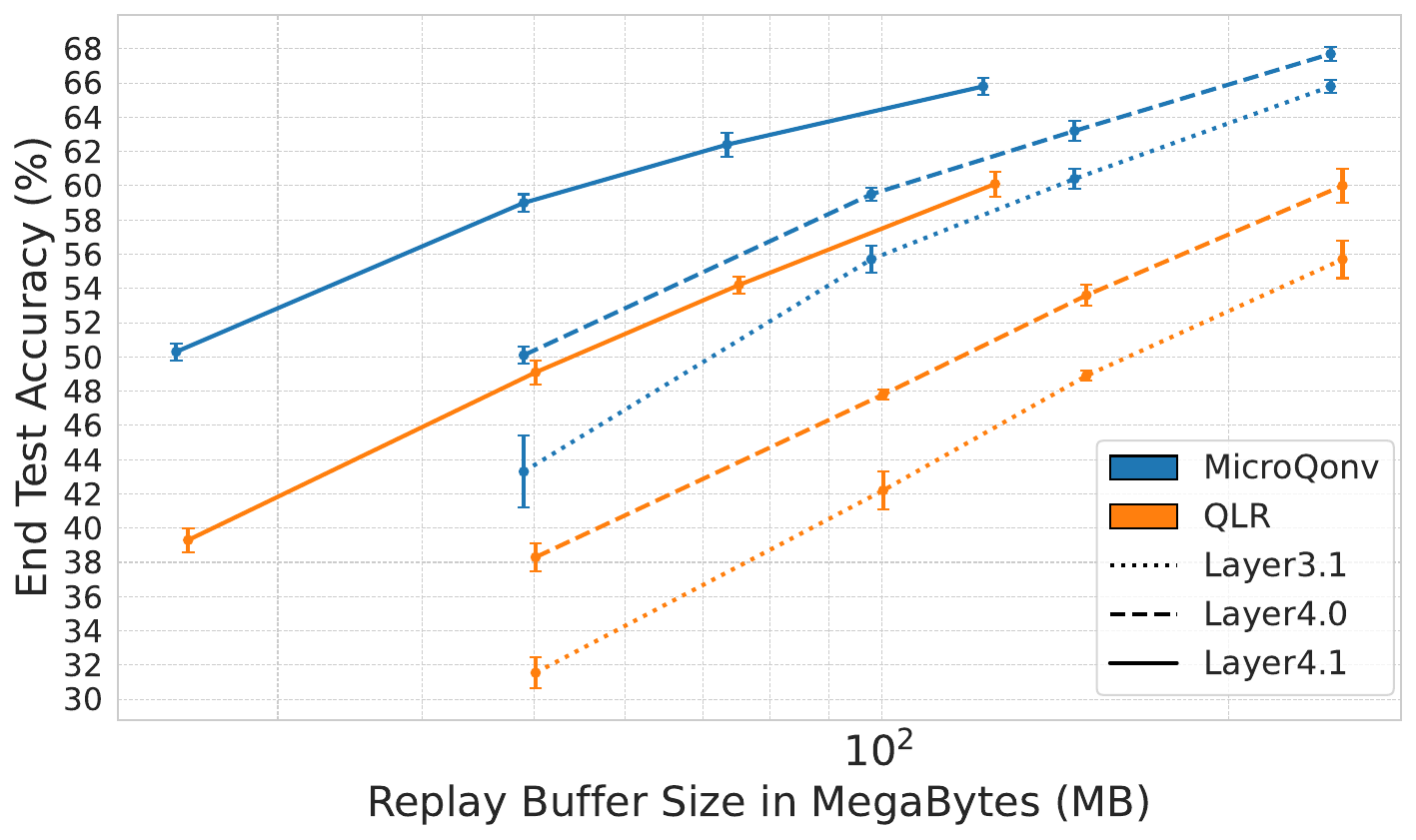}
    \caption{
        End-test accuracy vs replay buffer memory size.
    }
    \label{fig:pareto_memacc}
\end{figure}

To run our experiments, we combine the PyTorch-based PyCIL library~\cite{zhou_pycil_2023} with quantization methods, and we use the PyCIL hyperparameters for all methods. A ResNet18 model pretrained on ImageNet is quantized and full-precision accuracy retrieved through fine-tuning with Quantization-Aware Training~\cite{nagel_white_2021}: activations and weights are quantized in the tested formats, while gradients remain in full-precision. Each quantization method is used with the Hindsight technique~\cite{Fournarakis21Hindsight}, which estimates tensor statistics, such as the maximum and mean, from previous batches. This helps load each tensor only once, reducing memory movement. All methods quantize the first and last layers to 8 bits for weights, gradients, and activations, as a common practice to mitigate accuracy degradation. 
Only convolutional layers without parallel residual connections can be treated as LR layers. 
For our experiments, we select the first convolutional layer of blocks 3.1, 4.0, and 4.1.
We select samples for the replay buffer using the herding mechanism~\cite{Rebuffi2016iCaRLIC}, a widely used strategy.
Figure~\ref{fig:pareto_memacc} shows the test accuracy after training on the last task, in the CIL scenario on CIFAR-100. \MicroQonv\ accuracy is systematically and significantly above QLR across a range of layers and replay buffer sizes. Since 3.125 bits per activation pixel are saved instead of 8 bits, approximately $\times$2.5 more samples can be stored, which explains the accuracy delta. This would not be possible with other methods: Dacapo would save them using 32 bits per activation pixel, while Cuyckens \textit{et al.} extrapolation would save the expanded \imcol\ tensor, significantly reducing the number of samples that can be stored given a memory size.
%
%
As LR at layer 4.1 shows the best performance, we keep this configuration for the rest of the experiments. 

We now test \microqonv\ on the CIFAR-100, CORE50~\cite{lomonaco_core50}, and CUB-200 datasets. Each task has 10 classes for CIFAR-100, 5 for CORE50, and 40 for CUB-200.
Table~\ref{tab:CL_accuracy} lists the results at various memory sizes. \microqonv\ accuracy is always way above QLR, with test accuracy difference between +5.7\% and +11\% for CIFAR-100, +4.8\% and +9.4\% on CORE50, and +9.6\% and +16.7\% for CUB-200. The accuracy gap tends to be higher when there are fewer samples. This aligns with neural network scaling laws~\cite{sorscher22scalinglaws}: doubling the dataset size yields less accuracy improvement when the dataset is large than when it is small. We perform a final test to compare quantization performance on the CIFAR-100 dataset with a fixed memory budget of 2000 samples for all methods. The FP32 baseline achieves an accuracy of $48.83\pm.36\%$ while QLR attains $48.77\pm.23\%$ and \microqonv\ reaches $48.16\pm.66\%$, less than 1\% from baseline.


\begin{table}[tbp]
\centering
\caption{LR end test accuracy of ResNet18 for various LR memory sizes in MB and datasets.}
\label{tab:CL_accuracy}
\begin{tabular}{ccccc} 
\hline
  CIFAR-100& \textbf{25MB}  & \textbf{50MB}  & \textbf{75MB}  & \textbf{125MB}\\
\hline
    $\text{QLR}$\cite{ravaglia_tinyml_2021}  &  $39.3\pm.7$ &  $49.1\pm.7$ &  $54.2\pm.5$ &  $60.1\pm.7$   \\
    $\text{MicroQonv}$  &  $50.3\pm.5$ &  $59.0\pm.5$   &  $62.4\pm.7$ &  $65.8\pm.5$   \\
\hline
\hline
 CORE50 & \textbf{15MB}  & \textbf{25MB}  & \textbf{50MB}  & \textbf{75MB}\\
\hline
    $\text{QLR}$                &  $59.1\pm.5$ &  $64.0\pm.5$ &  $70.7\pm.8$ &  $73.2\pm1.1$   \\
    $\text{MicroQonv}$  &  $68.5\pm1.0$ &  $72.2\pm.4$   &  $75.5\pm.5$ &  $78.0\pm.8$   \\
\hline
\hline
 CUB200 & \textbf{10MB}  & \textbf{20MB}  & \textbf{30MB}  & \textbf{40MB}\\
\hline
    $\text{QLR}$     &  $26.6\pm.6$ &  $39.9.\pm.5$ &  $43.5\pm.7$  &  $50.2\pm.3$  \\
    $\text{MicroQonv}$   &  $43.3\pm.7$ &  $52.8\pm.3$  &  $57.1\pm.5$ &$59.8\pm.3$   \\
\hline
\end{tabular}
\end{table}

\section{Conclusion}\label{sec:conclusion}

\MicroQonv\ uses in-memory tensor reshaping for convolution kernels to simplify microscaling quantization and reduce quantization overhead by a factor of $\times2$ for weights and gradients, and up to $\times9$ for activations. By enabling the storage of a quantized activation tensor instead of its \imcol-expanded version, it reduces activation memory movement and storage by up to $\times7.53$ under standard convolution parameters. Validation with a DRAM simulator shows significant memory and energy gains on various models, including state-of-the-art object detection models: $\times3.5$ for YOLOV8nano and $\times2.2$ for YOLOV26nano. When applied to CL with an LR strategy, 4-bit \microqonv\ activations reduce memory usage and improve accuracy by +5.7\% to +11\% for the same memory budget. Future work includes building a hardware accelerator that leverages \microqonv\ memory layout and studying the relationships between other lossy compression techniques (e.g., vector quantization) and microscaling. We will release all code used in this work as open source upon publication.




\section*{Acknowledgment}
This work was supported by the PEPR IA - HOLIGRAIL project of the Agence Nationale de la Recherche, ANR-23-PEIA-0010.
Experiments presented in this paper were carried out using the Grid'5000 testbed, supported by a scientific interest group hosted by Inria and including CNRS, RENATER and several Universities as well as other organizations (see https://www.grid5000.fr)


\bibliographystyle{IEEEtran}
\bibliography{ref}




\end{document}